# From KBOs to Centaurs: The Thermal Connection

Gal Sarid
Department of Geophysics and Planetary Sciences, Tel-Aviv University, Tel-Aviv, Israel,
Tel Aviv University, P.O. Box 39040, Tel Aviv 69978, Israel,
galahead@post.tau.ac.il.

and

Dina Prialnik,
Department of Geophysics and Planetary Sciences, Tel-Aviv University, Tel-Aviv, Israel,
Tel Aviv University, P.O. Box 39040, Tel Aviv 69978, Israel,
dina@planet.tau.ac.il.

**Abstract**

We present results of thermal evolution calculations for objects originating in the Kuiper belt and transferring inwards, to the region of the outer planets. Kuiper belt objects (KBOs) are considered to be part of a reservoir that supplies the flux of small icy bodies, mainly Centaurs and Jupiter-family comets, to regions interior to the orbit of Neptune. We study the internal thermal evolution, for $\sim 10^8$ yr, of three typical KBOs and use the end state of the simulation as initial conditions for evolutionary calculations of two typical Centaurs. Some evolutionary trends can be identified for the KBOs, depending on key physical parameters, such as size and composition. The subsequent evolution in the Centaur region results in both specific features for each modeled object (mainly surface and sub-surface composition) and common characteristics of thermally evolved Centaurs.

**1. Introduction**

The outer solar system hosts a vast population of small icy bodies, considered to be primitive remnants from the planetary formation epoch. These are generally divided into sub-populations, according to their dynamical properties. One such primal division is between Kuiper belt objects (KBOs) and Centaurs. The former occupy the region of phase space with an aphelion distance grater than 30 AU, while the latter are on planet-crossing orbits between Jupiter and Neptune. These two populations are considered to be of the same origin, the Centaurs being objects that have been ejected and scattered inward. The dynamical evidence for this connection is strengthened by observations, as these populations appear to be similar in both colors and spectral characteristics. These relations between KBOs and Centaurs suggest that they share a common thermal and compositional history. Internal evolutionary models of representative Kuiper belt and Centaur objects could help in determining what primitive properties can still be found today and how icy bodies evolve in the outer Solar System region.

In the past decade, there has been a rapid increase in the number of discoveries of Centaurs: as of November 2008, the Minor Planet Center listed 237 Centaurs and scattered-disk objects (SDOs) as one group of objects (http://www.cfa.harvard.edu/iau/lists/Centaurs.html). Following a standard dynamical division between Centaurs and SDOs (Tiscareno & Malhotra 2003), in which the perihelion distance is interior or exterior to Neptune's orbit, a subset of ~120 objects belong to the family of Centaurs. Following another standard dynamical division (Jewitt 2004), in which the perihelia and semi-major axes satisfy $q > a_J$ and $a < a_N$, with $a_J = 5.2$ AU and $a_N = 30.2$ AU, a subset of 79 objects are defined as Centaurs. In Fig. 1 we show the distribution of all known Centaurs, SDOs and KBOs, according to their orbital elements. The distinction between the Centaur and scattered disk populations was made according to the second division mentioned.

The dynamical lifetimes of Centaurs are much shorter than the age of the solar system, so they must have a source in a more stable reservoir elsewhere in the outer solar system (Duncan et al. 2004). The prevailing view is that Centaurs are objects that have escaped from the region of the Kuiper belt beyond the orbit of Neptune, and represent a dynamical population intermediate between the relatively stable KBOs and the short-lived Jupiter-family comets. Once objects are trapped as Centaurs, the dynamical lifetimes are limited by strong gravitational perturbations from the giant planets to a median lifetime of $\sim 10^7$ yr (Tiscareno & Malhotra 2003). The fate of most Centaurs is to be scattered out of the solar system, but some survive and those that are scattered inwards, to orbits interior to that of Jupiter, are re-labeled as comets (Jewitt 2004).

Another recent evidence for the origin of Centaurs comes from broadband photometry and spectral observations of the brightest Centaurs. These objects show a wide diversity in color, from neutral/blue for 2060 Chiron, to ultra-red for 5145 Pholus (Barucci et al. 2005), much like the color diversity known for KBOs (Luu & Jewitt 2002). This color diversity, together with the evidence for spectral similarity between Centaurs and KBOs (Groussin et al. 2004, Barucci et al. 2004), are consistent with a common origin in the trans-Neptunian region.

The purpose of the present paper is to model the thermo-chemical evolution of a few typical KBOs, from an early epoch until a 'slowly-evolving' state is reached. By 'slowly-evolving' we mean that further thermal variations are small and the compositional and structural changes remain negligible. We then use the final structure as input for evolutionary calculations of Centaurs, spanning periods of time that are compatible with dynamical lifetimes. The end result should provide insight into the initial structure of comet nuclei of the Jupiter-family type (cf. Coradini et al. 2008).

Long-term evolutionary models of KBOs have been calculated by Choi et al. (2002), but simplifying assumptions were adopted there for the flow of gas through the porous nucleus that we avoid in this work. The significant result of that work was that highly volatile species – if present in the initial composition – would have been completely depleted in the very early stages of evolution. Evolution of KBOs, including release and flow of volatiles in the interior, was also studied by De Sanctis et al. (2001), but only for a relatively limited period of time, less than the decay time of the active short-lived radioactive species, and hence the full-scale effect of radiogenic heating could not be assessed. This study showed that volatiles flowing toward the interior may re-condense and form new ice. We thus attempt to improve on both these important studies, by carrying out long-term evolutionary simulations, including detailed calculations of gas flow through the porous interior and allowing for sublimation and re-condensation on the pore walls.

A brief description of the evolution code is given in the next section, results are presented in Section 3, and our main conclusions are summarized in Section 4.

## 2. Methodology and relevant physics

2.1 Description of the model

For the purpose of modeling the thermal evolution of KBOs and Centaurs we regard them as small spherical icy objects, similar to comet nuclei. A comet nucleus is generally portrayed as a porous aggregate of ices and solids (Weidenschilling 2004). As such, its structure can be modeled as an agglomeration of grains made of volatile ices and solids at some mixing ratio, with a wide spread in the size distribution of the components. The most general composition of a comet-like object is assumed to consist of water ice (amorphous and crystalline), water vapor, dust (grains of silicates and minerals), and other volatiles, which may be frozen, free or trapped in the amorphous water ice. The dust component may include radioactive elements in abundances typical of meteorites (Choi et al. 2002). Water ice is assumed to be initially amorphous, but will crystallize at a temperature-dependent rate (Schmitt et al. 1989) and gradually release the trapped volatiles.

The mass conservation equation for vapor is of the form

$$\frac{\partial \rho_v}{\partial t} + \nabla \cdot \mathbf{J}_v = q_v \ , \qquad (1)$$

where $\rho_v$ is the density, $\mathbf{J}_v$ is the flux and $q_v$ is the rate of gas release or absorption by sublimation/condensation (see Prialnik et al. 2004). The number of such equations to be solved is equal to the number of volatiles species considered.

The heat transfer equation is

$$\sum_\alpha \rho_\alpha \frac{\partial u_\alpha}{\partial t} - \nabla \cdot (K \nabla T) + \left( \sum_\alpha c_\alpha \mathbf{J}_\alpha \right) \cdot \nabla T = \lambda \rho_a H_{ac} - \sum_\alpha q_\alpha H_\alpha + \dot{Q} \ , \qquad (2)$$

where $\alpha$ is the index of each contributing species, $u$ is the internal energy, $K$ is the thermal conductivity, $c$ is the specific heat and the right-hand side (RHS) represents all the available energy sources (or sinks). These are bulk (body) sources and they are explicitly calculated in the heat transfer equation. This energy contribution is a result of heat release by crystallization (first term on the RHS), latent heat release of phase transitions (second term on the RHS) and heat released by radioactive decay (third term on RHS). Here, $\lambda$ is the crystallization rate, $\rho_a$ the density of amorphous ice, and $H_{ac}$ the heat released upon crystallization. $q_\alpha$ and $H_\alpha$ are the sublimation rate and heat of sublimation for each volatile, respectively. The sum over all present radionuclides and their heat due to radioactive decay is represented by $\dot{Q}$, where an exponential decay law was used with the properties of each element (see Choi et al. 2002 for a summary of the radionuclides' properties). Other energy sources, such as solar radiation and impact heating, which are external in their nature and depend on specific modeling, are included in the boundary conditions for the fluxes (Sarid et al. 2005).

All of the above considerations result in a set of time-dependent equations, which require additional assumptions about physical and chemical processes. These include heat conduction and gas flow in a porous medium, amorphous-crystalline transition, phase transitions (mainly for $H_2O$), radioactive heat production, growth or shrinkage of pores, due to sublimation/condensation. In addition, a number of initial parameters must be supplied, such as ice to dust ratio, volatile abundances and amount of occluded gas in the ice. The set of evolution equations are second-order in space, and hence require two boundary conditions: vanishing fluxes (heat and mass) at the center and energy balance at the surface. The code for the numerical modeling is described in detail by Prialnik (1992) and Sarid et al. (2005); a full discussion of the input physics may be found in Prialnik et al. (2004) and Sarid & Prialnik (2009). Table 1 lists the key input physics parameters for our thermal model. These parameters are set, at least initially, to be the same for all the models we calculated.

Self-gravity is usually neglected in small icy bodies, in comparison with the material strength, as even for very low compressive strength (~10 kPa), incompressibility is amply justified by the sizes and densities of comet nuclei (Prialnik et al. 2004). However, larger bodies ($\geq 100$ km) do not necessarily comply with this rule (Prialnik et al. 2008), as their sizes and densities are such that they become compressed under their own gravity, at least in most of their bulk. In our model we provide a hydrostatic pressure and density distribution, assuming an equation of state (EOS) for the solid matrix (Sarid & Prialnik 2009). We take this to be the Birch-Murnaghan equation of state, which is one of the simplest and most commonly used in planetary science (Poirier 2000). Prialnik & Sarid (2009) discuss self-gravity and varied EOSs and show that self-gravity can be quite safely ignored below radii of ~30 km, but has to be taken into account for radii larger than ~100 km, even if we consider different values of compressive strengths, bulk densities and limiting spin periods.

2.2 Choice of sample objects and initial parameters

In order to follow the thermal evolution of an object from the region of the Kuiper belt to the neighborhood of the outer planets, we need to choose the parameters of the bodies to be modeled. The objects presented here, both KBOs and Centaurs, are among the best observed objects of each group. This means that although our parameter space of physical characteristics is still large, these objects represent a group for which we know at least some of the basic attributes.

The parameters for the selected KBOs are listed in Table 2. Knowing the estimate for the bulk density, we can constrain the proportions of ice, dust (solids) and voids (porosity), using the relation

$$1-\Psi = \frac{\rho}{1+\Upsilon_{d/i}}\left(\frac{1}{\rho_i}+\frac{\Upsilon_{d/i}}{\rho_d}\right) \; , \qquad (3)$$

where $\Psi$ is the porosity (either initial or a given point), $\Upsilon_{d/i}$ is dust to ice ratio of mass fractions and ($\rho, \rho_i, \rho_d$) are the bulk density and specific solid density of ice and dust, respectively. We have chosen the initial parameters so as to take in consideration the estimated mass and compositions of these objects. These objects are part of a parameter study of medium to large KBOs that will be presented elsewhere (Sarid & Prialnik 2009). Briefly stated, the values in Table 2 sample several regions in the size-density-composition parameter space.

In order to have a common ground to compare the different results of the evolution runs, we have to take structural and compositional parameters to be similar. The most important ones for our numerical simulations, are listed in Table 1 and Table 3, where the initial abundances of the radioactive species and the volatiles appear. For the composition, other than water ice, we take the three most common compounds in planetary environments - $CO$, $CO_2$ and $HCN$ (Bergin et al. 2007). These are also among the most abundant cometary volatiles observed (Bockelée-Morvan et al. 2004).

As mentioned previously, we assume that water ice is in amorphous phase initially and the volatile compounds are trapped as occluded gas in the ice, to be released upon crystallization (Bar-Nun et al. 1987). Indirect support for the original amorphous state of water ice, in the interior of cometary bodies, comes from observations of highly volatile species, such as $CO$, in the comae of comets (Bockelée-Morvan et al. 2004). The equilibrium temperature at distances corresponding to the Kuiper belt region (between ~30-60 K) are above the condensation temperature of highly volatile species, but low enough for water ice to be amorphous. For $CO_2$ and $HCN$, which are moderately volatile, the initial ice phase was considered negligible due to the heat deposition during the accretion phase of KBOs, which may be sufficient for preventing condensation of these volatiles, but not so strong as to drive the temperature above the crystallization threshold (Shchuko et al. 2006). Thus water ice remains amorphous for the initial configuration of our thermal model (cf. McKinnon 2002).

Considering the long term evolution of objects in the trans-Neptunian region, an energy source derived from the decay of radioactive isotopes must be taken into account (Choi et al. 2002, McKinnon et al. 2008). We take the initial mass fractions of radioactive species from meteoritic studies, attenuated by a factor of $5 \times 10^6$ yr of accretion time. Objects with radii larger than several 10-100 km are formed from smaller icy planetesimals. These smaller objects are inefficient in storing internally produced heat (Prialnik & Podolak 1999) and so radioactive decay causes a decrease in the abundances of radionuclides without much thermal alteration of the interior, prior to their inclusion into larger conglomerates. Once a large enough object is formed, it can retain its internally produced heat. Thus, the effective initial abundances of radioactive isotopes, for the purposes of thermal modeling, are those obtained at the final stages of accretion. Calculations of planetesimals formation predict a few $10^7$ yr of formation timescale, for objects with radii larger than 100 km and formation regions at 30-50 AU (Kenyon 2002). However, the formation process should be more rapid if a more massive solar nebula, or more compact formation zone, is considered, as some models suggest (Desch 2007, Kenyon et al. 2008). Thus, to account for these, and other, uncertainties we have taken a shorter accretion time for large KBOs. It should be noted that this is an approximation, as it would be strictly correct only if the build-up of such objects was very slow for the duration of the accretion time and very rapid afterwards (Merk & Prialnik 2003). Since our time span for the thermal simulation runs is $10^8$ yr, only the short-lived radioactive species, with half-lives around $10^6$ yr, contribute appreciably to the internal heat balance. The long-lived species are however included in the calculations, for completeness.

The time scale for thermal evolution simulations is limited by computational constraints. Thus, we have chosen it so as to enable a sufficiently long evolution term for each object (i.e., to track long term trends) and to be in agreement with collisional and diffusion time scales in the Kuiper belt, as derived from dynamical studies (Durda & Stern 2000, Morbidelli 2008).

The sample Centaur objects chosen are 10199 Chariklo and 8405 Asbolus. They are among the best-characterized Centaur objects and their surface features have been measured and classified by several studies (e.g., Fernandez et al. 2002, Dotto et al. 2003, Barucci et al. 2004). From a dynamical point of view, these two objects probably represent distinct sub-groups within the Centaur population (Horner et al. 2004). The measured parameters for each of the Centaur objects are listed in Table 4. Asbolus has a high-eccentricity orbit, with a perihelion distance of 6.83 AU, while Chariklo's low-eccentricity orbit has a perihelion distance of 13.07 AU. This means that Asbolus has a maximum input energy from solar radiation greater than Chariklo's. If we calculate the average incoming power (from solar radiation) per unit area, we see that Asbolus gains a few more percent than Chariklo. Another consideration for the thermal simulation of these objects is their dynamical lifetime. Since the Centaurs are considered a transient population, most objects occupy stable orbits for a relatively short period of time. However, this does not necessarily hold true for every specific object and there is a wide dispersion in calculated lifetimes (Tiscareno & Malhotra 2003).

Dynamical simulations of large sets of test particles, with initial orbital elements slightly varied around their known mean values, can yield estimates for individual objects' dynamical lifetime. We report briefly here the results of some dynamical simulations, with regards to the mean dynamical lifetime of the two Centaur objects, Asbolus and Chariklo. We ran several simulations, as described above, for initial populations of 1000 test particles, having initial orbital elements randomly chosen around their currently measured values (see Table 4). For the calculations we used the regularized symplectic integrator, as it is implemented in the SWIFT integration package (Levison & Duncan 1994; Morbidelli 2002). We tracked the variations in the orbital elements of these mass-less test particles, influenced by the gravitational perturbations of all the giant planets. A particle was removed from the simulation in the following situations: evolution of orbit to a heliocentric distance smaller than 5.2 AU (Jupiter's orbit), evolution of orbit to a heliocentric distance greater than 30.2 AU (Neptune's orbit), heliocentric distance reaching a distance greater than 100 AU, or collision onto a planet (particle-planet distance smaller than the planet's radius). The time step was constrained to be 1 yr and the overall time span of the simulations was $10^7$. We define as the dynamical lifetime of an object the time during the simulation when 90% of its cloned test particles have been removed. Our results, that will be expanded elsewhere, indicate that for Asbolus, the dynamical lifetime estimate is $9.4 \times 10^5$ yr, while Chariklo exhibits far greater stability with a dynamical lifetime estimate of at least $10^7$ yr. For the purpose of our thermal evolution simulations we chose the time span for calculation as a fraction of the dynamical lifetime estimates. This results in $\sim 3\times 10^4$ yr and $\sim 3\times 10^6$ yr, respectively.

In order to examine the influence of the dynamical diversity on their thermo-chemical evolution, the initial composition of Asbolus and Chariklo was taken to be the same. Table 5 lists the initial structure and composition of the Centaur models. Since we wish to explore the evolution of cometary-like objects, as they transfer from the Kuiper belt to the region of the outer planets, the initial conditions for the Centaur simulations were derived from the KBOs simulations, described above (more details are given in the next section).

## 3. Results of numerical simulations

3.1 Evolution in the Kuiper belt

We follow the internal evolution of three KBOs, (50000) Quaoar, 1992 QB1 and 1998 WW31, from an initially homogeneous configuration. Each object was evolved up to $10^8$ yr, through consideration of all available energy sources. Since energy sources (or sinks) resulting from latent heat release (or absorption) during phase transitions, including the amorphous to crystalline transition, are temperature-dependent, some other energy source is required to increase the temperature above the appropriate threshold. This source is the heat released by the radioactive decay of $^{26}$Al. Because of its short half-life ($<10^6$ yr) and hence higher heat generation rate, it dominates over the long-lived isotopes and the short-lived $^{60}$Fe sources.

Fig. 2 shows temperature profiles for each of the three KBOs at several times during the evolution. The build-up of internal heat, around the time corresponding to the half-life of the short lived radionuclides, is evident in all of the objects. However, only Quaoar and QB1, with $R \geq 100$ km, retain relatively high temperatures ($T > 200$ K) in their interiors for long durations. At the end of the simulations ($\sim 10^8$ yr), a slowly-evolving state is reached in the interiors of the all objects. We confirm previously derived results about the depletion of highly-volatile species at an early stage of the objects' evolution (Choi et al. 2002). These species are represented by *CO* in our initial compositions and neither ice nor gas phase is present at almost any stage of the calculation, except for what survives as occluded gas in the amorphous ice. Once the amorphous ice crystallizes, the super-volatile gas flows through the pores to the surface and escapes. We note that the temperature distributions at earlier times can exhibit a 'wavy' behavior, clearly seen down to a relative depth of ~0.4-0.5, for Quaoar and 1998 WW31. This is due to the ongoing thermal processing in the interior, which induces changes in structure and composition. When a layer enriched in volatile ices, because of previous thermal alteration of the deeper layers and subsequent outflow and condensation of volatile gases, experiences a rise in temperature it may sublimate. This process of sublimation and escape of gas, towards the colder outer layers, reduces the local temperature. As gas sublimates and condenses throughout the bulk, from the interior outwards, following the temperature evolution, the porosity also changes. A change in porosity is caused by the sublimation or condensation of volatiles onto the pore walls. This changes the local pore-size distribution. As the thermal and gas flow parameters depend on porosity and local temperature, we get an elaborate interplay between the sublimation/condensation of various volatiles, crystallization of amorphous ice, which is a continuous process that releases energy and the temperature and pressure gradients. As the object evolves, it becomes depleted in volatiles and amorphous ice, and its radioactive energy reservoir is attenuated. Thus, thermal and compositional variations are quenched and the temperature distribution becomes smoother. This is shown by the solid curves, corresponding to evolution after $\sim 10^8$, in Fig. 2 a-c.

Due to its large size, Quaoar retains a temperature close to 200 K in its interior for most of its evolution. The relatively high temperature attained throughout most of the volume of Quaoar, from a very early state to the end, results in almost complete crystallization, with layers of amorphous ice surviving only close to the surface. These "sub-surface" layers comprise only less than 5% of the radius, but still extend to ~30 km of depth. At this depth the amorphous ice (and its occluded volatiles) is well-protected from the effects of space weathering (irradiation by high-energy particles) and impact gardening (collision of small-scale bodies onto the surface of a celestial object). It is noteworthy that, in addition to the amorphous ice surviving at this depth, there is also a layer of re-frozen volatile ice. This is the result of gas flow from the heated interior, where the amorphous to crystalline water ice transition occurs, to the outer cooler layers. The gas, comprised of $CO_2$ and $HCN$ molecules, flows through the porous solid matrix and eventually condenses. This is shown in Fig. 3, which is the result of an extended evolution of Quaoar, up to $\sim 10^9$ yr. Appreciable amounts of volatile ice layers (with respect to the initial volatile abundances) survive for over $5 \times 10^8$ yr at depths down to ~120 km.

The sublimation and condensation of various volatiles and the subsequent increase or decreases of gas flow determine the effective gas pressure. This build-up of pressure may occur in regions where the local porosity changes and may induce conditions of blow-off, or fluidized bed flow, if the gas pressure inside the pores exceeds the local hydrostatic pressure. By examining the gas and hydrostatic pressure distribution throughout the interior of the objects, we conclude that at no point do we have appropriate conditions for the situations mentioned above. This is because the regions where hydrostatic pressure is low (outer layers) are also the ones with very cold conditions, so gas pressure is negligible. In the interior, where there may develop conditions for moderate gas flow (at least at early times, when temperatures are higher and the volatiles are no yet depleted), the hydrostatic pressure is high.

3.2 Subsequent evolution in the region of the outer planets

In order to follow the evolution of Centaurs, as objects having early thermal histories in the Kuiper belt, we assume a physical connection between the objects simulated. Two scenarios are considered: (I) a relatively small fragment of a much larger, thermally processed, KBO evolves dynamically to become a Centaur; (II) a smaller KBO evolves thermally beyond Neptune's orbit and then diffuses inwards to become a Centaur. In the first case, we assume that the collision that produced the fragment imparted mostly kinetic energy. Thus, the fragment quickly becomes a Centaur and the heat deposition during the collision is too small to induce thermal alteration of the fragment's interior. In the second case, no external effects are taken into account and the final state of the KBO simulation becomes the initial state of the Centaur simulation. According to the current understanding of the collisional history of the Kuiper belt (Kenyon et al. 2008), the first scenario may be applicable to smaller objects, such as Asbolus, while the latter scenario could represent the larger Centaur objects, such as Chariklo.

We thus assume that Asbolus – and similar objects – were initially debris ejected, after a catastrophic collision, by a KBO resembling Quaoar. Accordingly, we take the final state of the upper ~20% of Quaoar's volume resulting from the extended simulation mentioned in the previous subsection, and use it to construct homogeneous initial models for the object's bulk. As the initial bulk density we take the weighted average of the density profile in the outer layers of the evolved Quaoar model. The initial temperature is chosen as the smaller between the equilibrium temperature at perihelion in the Centaur's orbit and the average temperature of the upper layers of the evolved Quaoar model. This amounts to 65 K for Asbolus. The initial composition is taken as a weighted average of the mass fractions in the upper layers, as listed in Table 5. Since the initial state of the interior of Asbolus is derived from an evolved model, it is composed of both amorphous and crystalline water ice. $CO_2$ and HCN are present both in gas and ice phase. CO is only included as trapped gas in the small fraction of amorphous ice. We assume that there is no contribution of heat from radioactive isotopes, as the short-lived ones have already decayed. The long-lived radionuclides are still of negligible importance and the duration of the evolution ($\sim 3.3 \times 10^4$ yr) is not long enough to enable their on-going decay to accumulate any significant heat.

For Chariklo, we assume that a KBO resembling 1992 QB1, which has similar size and albedo as Chariklo, diffused inwards of Neptune's orbit, after a thermal evolution period of $\sim 10^8$ yr in the Kuiper belt zone. Thus, the initial structure is identical to that of the evolved model of QB1 and the initial temperature distribution is that of the final orbit calculated in the Kuiper belt. Table 5 lists the initial parameters for the model of Chariklo, where the previous KBO thermal evolution is evident in the mass fractions of amorphous and crystalline water ice and the presence of non-negligible abundances of volatile ices.

The internal evolution of Chariklo is shown in Fig. 4 (temperature distribution) and Fig. 5 (structure and composition distribution). The surface very rapidly cools to the ambient temperature of ~70 K and remains so, as the cooling and heating rate of the surface are similar. From that point on, the only change in temperature is the slow heating of the sub-surface depths corresponding to the penetration depth of heat from solar radiation. At the end of the simulation, most of its interior retained the initial temperature distribution and the outer few 100 m are at equilibrium with the surroundings. The minor increase of temperature at a depth of 1-10 km is not enough to trigger any appreciable sublimation or crystallization. Considering the internal structure and composition, we note that there is almost no change from the initial state. Density and porosity distributions remain almost constant in time and the amorphous ice abundance is slightly diminished, as it crystallizes at a very slow rate, as can be seen in the bottom-left panel of Fig. 5. Volatile ices remain at the same depth, just under the skin depth (heat penetration depth), with slight variations. These variations are slow and minor, because the corresponding temperature at this depth (3-10 km) is 80-90 K, which is the range of sublimation temperatures for CO$_2$ and HCN (Meech & Svoren 2004).

The internal evolution of Asbolus is shown in Fig. 6 (temperature distribution) and Fig. 7(structure and composition distribution). The surface is heated by the successive passages at less than 7 AU from the Sun. At this distance the equilibrium temperature is ~110 K and thermal processing occurs, as the heat absorbed near-perihelion is conducted inwards. The progressing heat front is evident in Fig. 6, as there is clear heating at all depths, down to ~5 km. After reaching its highest temperatures at its orbital skin depth (~25 m), Asbolus continues to heat up slowly. The boundary region between the colder interior and the hotter outer layers reaches ~3 km at $\sim 10^4$ yr. From this point on, the heat front progresses much more slowly inwards. This is due to the change in abundances of amorphous and crystalline water ice and volatile ices resulting from the earlier thermal processing. Only slight variations in density and porosity occur during the evolution. However, the penetrating heat front produces more pronounced variations in abundances of amorphous water ice (up to ~5%) and volatile ices (up to ~80%). In the profiles shown in Fig. 7 it is also clear that a depth of ~1 km constitutes a boundary region between thermally-processed and relatively pristine material (relative to the initial state).

## 4. Conclusions

In this paper we presented results of evolution sequences for models of a few typical KBOs, from an initial homogeneous state to a slowly-evolving state, during which they were thermally processed in the interior, to various degrees, until further variations in the interior were small and the object simply went through a process of cooling. The objects considered varied in size, initial bulk density and ice to dust ratio. However, the major constituents of their initial composition (other than water), as well as their sources of internally produced heat by radioactive decay, were kept the same.

Simply by looking at the peak temperatures attained during the simulations, we can relate the degree of thermal processing to size. Objects of sizes between 10 and 100 km reach the slowly-evolving state relatively quickly ($<10^8$ yr) and cool efficiently. Larger objects require more time and are thermally processed to a greater extent. These objects should be able to maintain a large fraction of their inner volumes (about 1/3 of their size) at almost 200 K, for the duration of the evolution. We must keep in mind, however, that the thermal behavior of our sample objects is a direct consequence of the initial radioactive isotope abundances, as solar radiation is negligible at heliocentric distances characteristic of the Kuiper belt (Choi et al. 2002; Coradini et al. 2008).

We also find that the final structure of bodies with $R > 100$ km is stratified in its composition (see also De Sanctis et al. 2000) and may exhibit "pockets" of volatile ices, such as $CO_2$ and $HCN$, at depths of several km to several tens of km beneath the surface. These depths are accessible to dynamical processes such as impacts that may result in ejection of debris and depression features of the surface and sub-surface layers, exposing fresh material and leading to slow sublimation of volatile compounds.

After a slowly-evolving state was attained, we used the final structure as input for evolutionary calculations of Centaurs. By applying the same modeling procedure but for different initial configurations, which represent different scenarios for the origin of Centaur objects, and different orbital behavior it is possible to distinguish between degrees of thermo-chemical processing based on dynamical features and origin. A general common result for the Centaurs simulations is that regardless of the specific thermal behavior during the evolution, a "quiet" configuration, with no further variations of internal properties, is reached throughout most of the object's volume (mid-layers to the deep interior) on a time scale of a few percent of the dynamical lifetime. Since we started the calculations from an already processed internal state, the dependence of the time scale required to reach this configuration on the internal composition is probably weak, if not negligible.

By contrast, the surface and sub-surface layers, which may span up to ~10% of the radius in depth, may experience continuous heating, as in the case of 8405 Asbolus. This depends on the perihelion distance and eccentricity, as objects that come closer to the Sun absorb more heat and objects with larger eccentricities experience a greater difference between maximal and minimal ambient temperature at the surface. Composition differences between different Centaurs, as revealed in photometry and spectroscopy of their surfaces, could be the result of relatively recent thermal evolution, rather then a consequence of different initial states.

Thus, we can say that a wide range of specific evolution outcomes is possible for KBOs, depending on size, location and initial structure and composition. However, by carefully mapping the parameter space of their physical characteristics, we can identify some evolution trends and the most important factors influencing them. Continued evolution in the Centaurs region results in a two-component division of the objects bulk - most of the volume reaches conditions for a slowly-evolving state on relatively short evolution time scales for all objects, while the surface and sub-surface layers may exhibit diverse features in composition and structure.

## Acknowledgements

This work was supported by grant 859-25.7/2005 of the German-Israeli Foundation for Scientific Research and Development. We would like to thank D. Jewitt and an anonymous reviewer for very helpful comments and suggestions that improved the quality and clarity of the manuscript.

**Tables**

Table 1. Common input physics parameters for all models.

| Parameter | Value | Units |
|---|---|---|
| Specific $H_2O$ ice density | 0.917 | g cm$^{-3}$ |
| Specific dust density | 3.25 | g cm$^{-3}$ |
| Heat capacity of ice | $7.49 \times 10^4 T + 9 \times 10^5$ | erg g$^{-1}$ K$^{-1}$ |
| Heat capacity of dust | $1.3 \times 10^7$ | erg g$^{-1}$ K$^{-1}$ |
| Conductivity of C-Ice | $5.67 \times 10^7 / T$ | erg cm$^{-1}$ s$^{-1}$ K$^{-1}$ |
| Conductivity of dust | $10^6$ | erg cm$^{-1}$ s$^{-1}$ K$^{-1}$ |
| Diffusivity of A-Ice | $3 \times 10^{-3}$ | cm$^2$ s$^{-1}$ |
| Pore-size distribution [1] | -3.5 | |
| Pore-size range [2] | $10^4$ | |

[1] Exponent for initial power-law distribution of pore sizes, as in comets (Sarid et al. 2005). [2] Ratio of maximal to minimal pore radius, as in comets (Sarid et al. 2005).

Table 2. Model parameters for each KBO.

| | 1998 WW31 | 1992 QB1 | 50000 Quaoar |
|---|---|---|---|
| a [AU] [1] | 39.13 | 43.73 | 43.50 |
| e [1] | 0.270 | 0.064 | 0.035 |
| R [km] | 65 | 100 | 630 |
| $\rho$ [kg/m$^3$] | 1500 | 750 | 1500 |
| A [2] | 0.07 | 0.04 | 0.10 |
| Ref. [3] | Veillet et al. 2002 | Jewitt et al. 1992 | Brown & Trujillo 2004 Luu & Jewitt 2002 |
| $\Upsilon_{d/i}$ init. [4] | 4 | 1.2 | 2.3 |
| $\Psi_{init.}$ [5] | 0.3 | 0.5 | 0.2-0.3 |

[1] Orbital parameters from MPC database. [2] Albedo. [3] Reference for the radius, density and albedo estimates. [4] Initial dust to ice ratio. [5] Initial bulk porosity.

Table 3. Composition parameters for KBO models.

| | |
|---|---|
| $X_{r,\ init.}$ (short-lived) [2] | $^{26}$Al: $4.79\times10^{-9}$ |
| | $^{60}$Fe: $2.81\times10^{-8}$ |
| $X_{r,\ init.}$ (long-lived) [2] | $^{40}$K: $8.78\times10^{-7}$ |
| | $^{232}$Th: $4.39\times10^{-8}$ |
| | $^{238}$U: $1.76\times10^{-8}$ |
| | $^{235}$U: $5.02\times10^{-9}$ |
| $X_{v,\ init.}$ [3] | CO      CO$_2$     HCN |
| | 0.005    0.005    0.0025 |

[1] Initial value for the power law exponent of the pore size distribution (Sarid et al. 2005).
[2] Initial mass fraction of radioactive species. [3] Initial fractions of volatile species occluded in the amorphous ice.

Table 4. Model parameters for each Centaur.

| | 10199 Chariklo | 8405 Asbolus |
|---|---|---|
| a [AU] [1] | 15.795 | 18.096 |
| e [1] | 0.172 | 0.622 |
| i [deg] [1] | 23.4 | 17.6 |
| R [km] [2] | 140 | 33 |
| A [2] | 0.05 | 0.12 |

[1] Orbital parameters from MPC database. [2] Radius and Albedo estimates From Dotto et al. 2003 (Chariklo) and Fernandez et al. 2002 (Asbolus).

Table 5. Initial parameters for the Centaur models.

| | 8405 Asbolus | | | 10199 Chariklo | | |
|---|---|---|---|---|---|---|
| $\rho$ [kg/m$^3$] [1] | 1100 | | | 0.75 | | |
| $\Upsilon_{d/i}$ init. [2] | 2.33 | | | 1.2 (same as 1992 QB1) | | |
| $X_{am.}$ ; $X_{cry.}$ [3] | 0.065 ; 0.23 | | | 0.2 ; 0.1 | | |
| $X_{v,\ init.}$ [4] | CO | CO$_2$ | HCN | CO | CO$_2$ | HCN |
| | 0.005 | 0.005 | 0.0025 | 0.005 | 0.005 | 0.0025 |
| $X_{ice,\ init.}$ [5] | CO$_2$ | HCN | | CO$_2$ | HCN | |
| | 0.004 | 0.001 | | 0.003 | 0.0014 | |

[1] Average bulk density. [2] Initial dust to ice ratio. [3] Initial mass fractions of amorphous and crystalline ices. [4] Initial fractions of volatiles occluded in the amorphous ice. [5] Initial mass fractions of volatile ices.

**Figures**

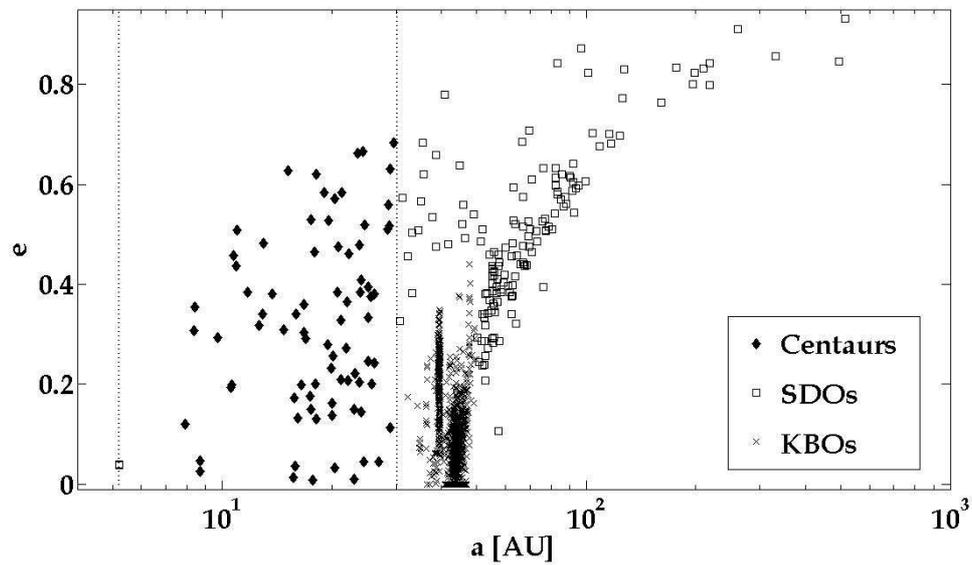

Fig. 1. Distribution of Centaurs, SDOs and KBOs on the orbital elements space of semi-major axis and eccentricity. The two dotted horizontal lines represent the semi-major axis of Jupiter (5.2 AU) and Neptune (30.2 AU). The data here is taken from the Minor Planet Center lists as of November 2008.

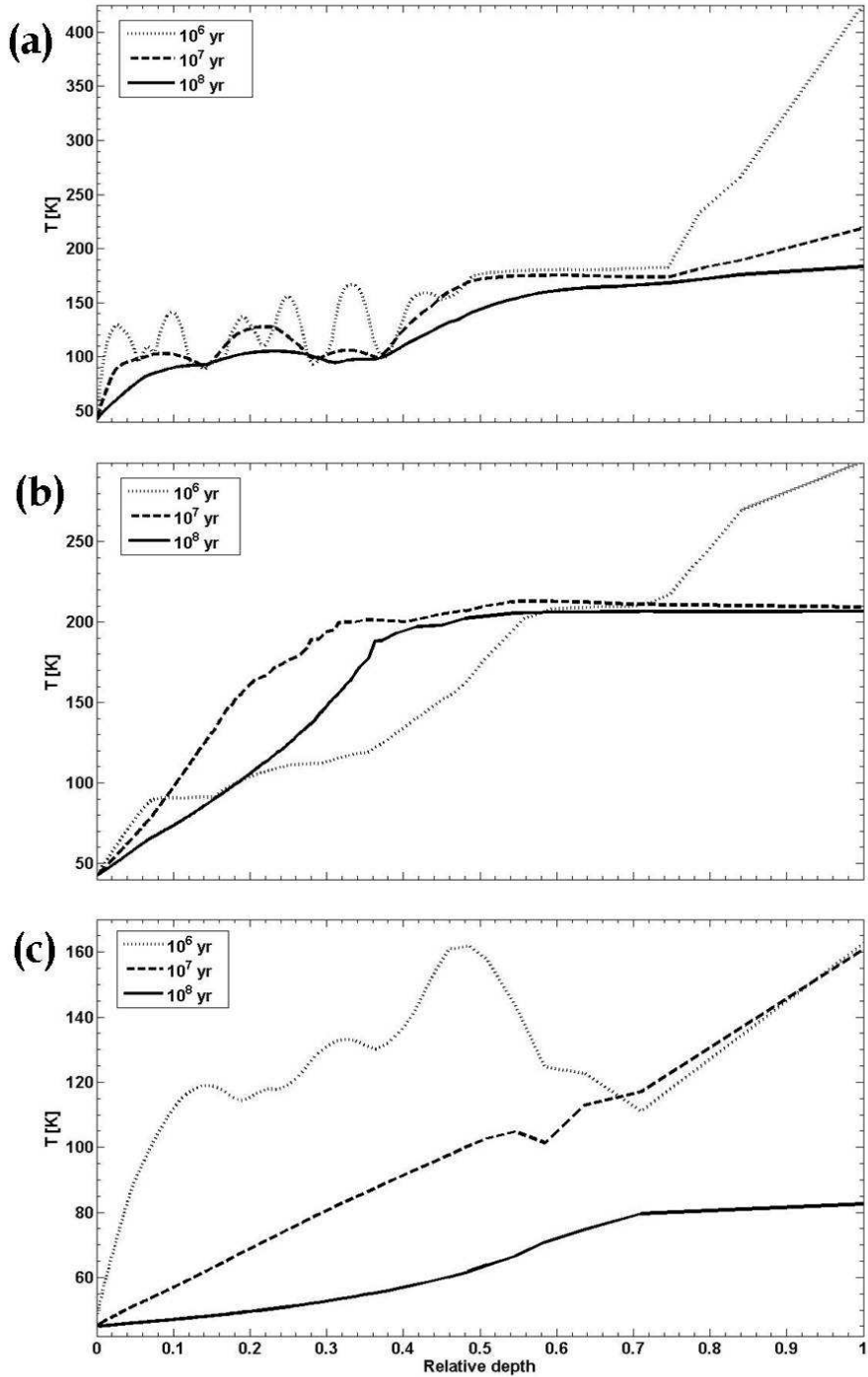

Fig. 2. Temperature distribution for 50000 Quaoar (a), 1992 QB1 (b) and 1998 WW31 (c). Depth scale is normalized to the object's radius (see Table 2), with 0 being the surface and 1 the center. Temperatures are presented for various times during the evolution: $10^6$ yr (dotted), $10^7$ yr (dashed) and $10^8$ yr (solid). Maximal temperatures attained during the evolution are 420 K, 300 K and 190 K for Quaoar, QB1 and WW31, respectively.

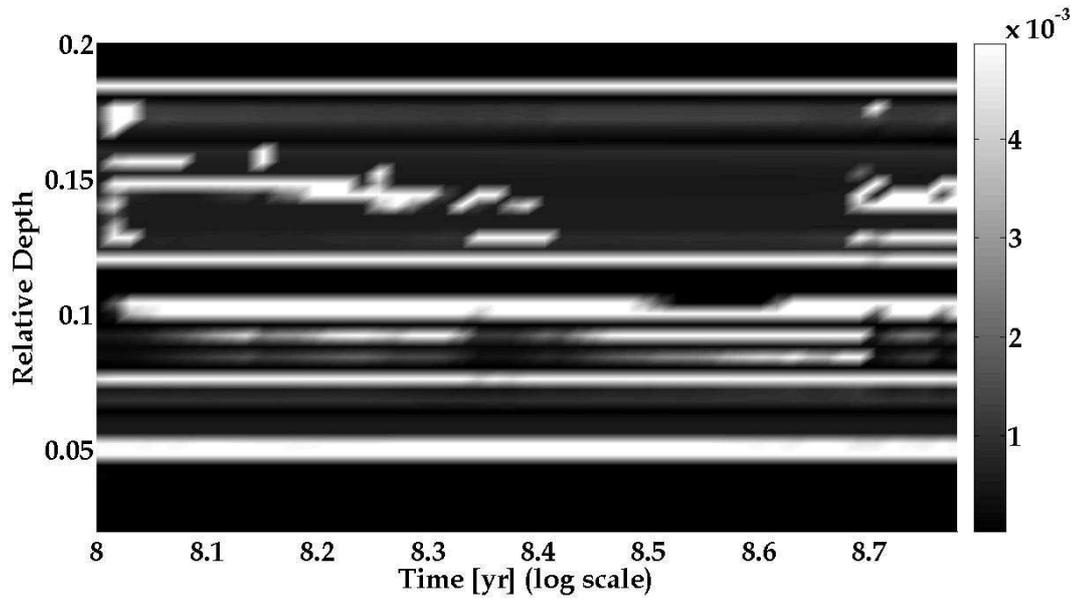

Fig. 3. Profile of the volatile ice mass fraction ($CO_2$ and $HCN$) for 50000 Quaoar. Time is given on a logarithmic scale and the depth scale is normalized to the Quaoar's radius, with 0 being the surface. Displayed here are the last $\sim 5.3 \times 10^8$ yr of evolution in the top 20% of the body's depth.

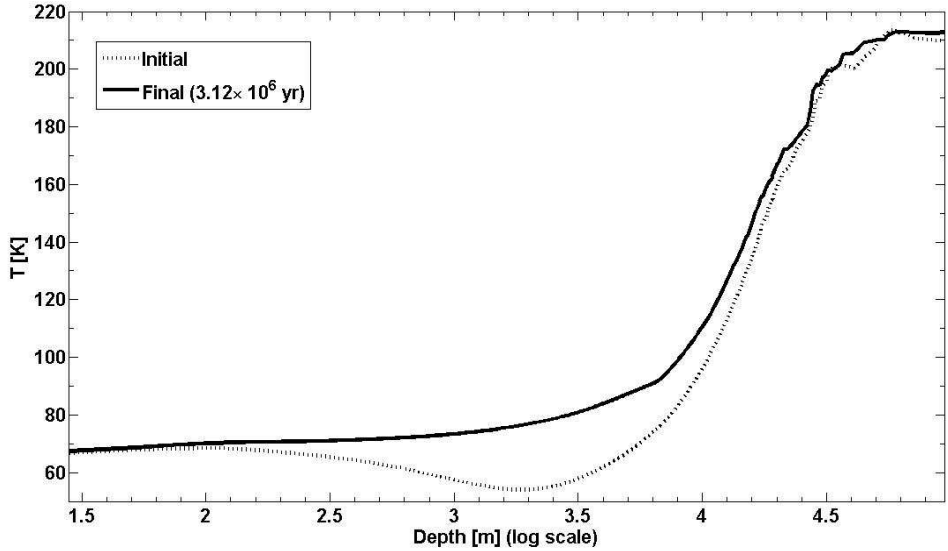

Fig. 4. Temperature profiles for 10199 Chariklo. Displayed for comparison are initial and final ($3.12 \times 10^6$) evolution states of the simulation. Depth is given on a logarithmic scale, from left (surface) to right (center). Note that the temperature is almost constant in time, except for layers at a depth of ~1-10 km below the surface, where heat builds-up around the orbital skin depth.

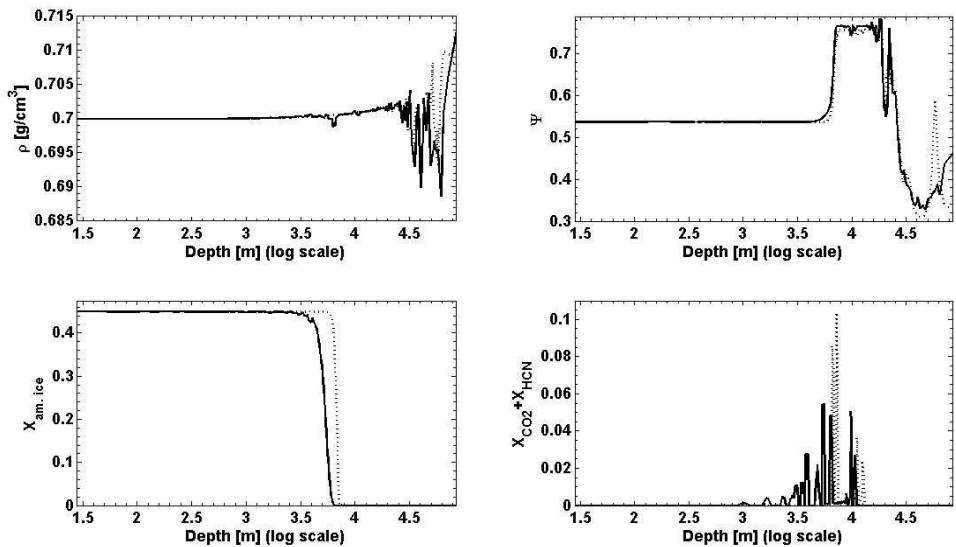

Fig. 5. Profiles of density (top left), porosity (top right), amorphous ice (bottom left) and volatile ice (bottom right) mass fractions for 10199 Chariklo. Depth is given on a logarithmic scale, from left (surface) to right (center). Displayed for comparison are the initial non-homogeneous state (dotted) and final state of the simulation (solid) at $3.12 \times 10^6$ yr.

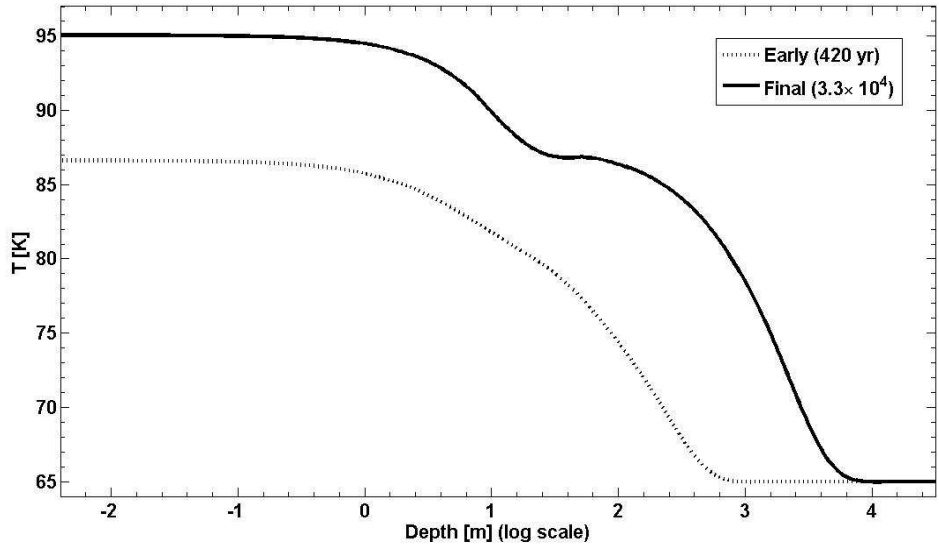

Fig. 6. Temperature profiles for 8405 Asbolus. Displayed for comparison are early and final evolution times in the simulation: 420 yr (dotted) and $3.3\times10^4$ yr (solid). Depth is given on a logarithmic scale, from left (surface) to right (center). Note the larger temperature variations, in comparison to Fig. 4.

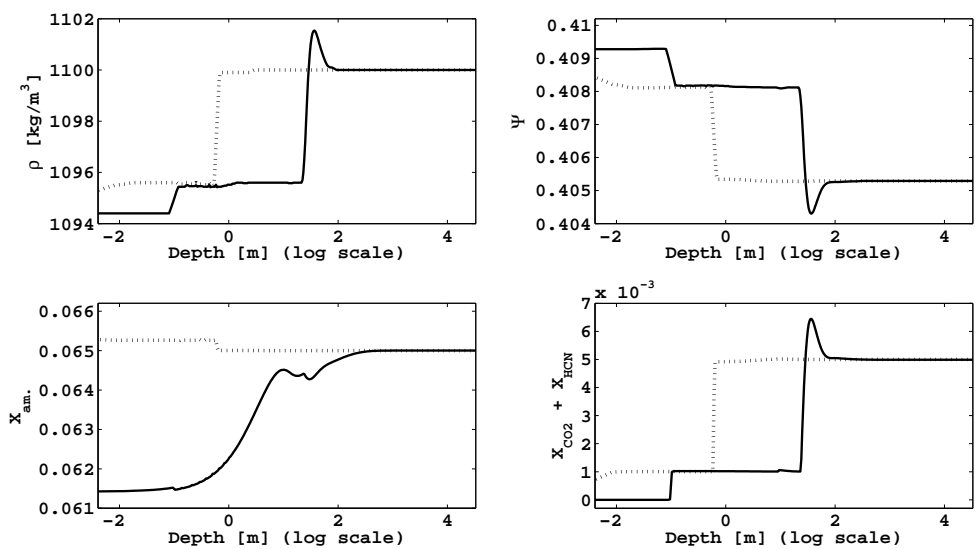

Fig. 7. Profiles of density (top left), porosity (top right), amorphous ice (bottom left) and volatile ice (bottom right) mass fractions for 8405 Asbolus. Depth is given on a logarithmic scale, from left (surface) to right (center). Displayed for comparison are early and final evolution times in the simulation: 420 yr (dotted) and $3.3\times10^4$ yr (solid).